\documentclass[twocolumn,showpacs,aps,prl]{revtex4-1}

\usepackage{amsmath,amsfonts,amssymb}
\usepackage{wrapfig}
\usepackage{graphicx}
\usepackage{bbm}
\usepackage{color}
\newcommand\bra[1]{\left\langle{#1}\right|}
\newcommand\ket[1]{\left|{#1}\right\rangle}
\usepackage{url}

\begin{document}
\title{Quantum process capability}

\author{Chung-Cheng Kuo$^{1,2}$}
\author{Shih-Hsuan Chen$^{1,2}$}
\author{Wei-Ting Lee$^{1,2}$}
\author{Hung-Ming Chen$^{1,2}$}
\author{He Lu$^{3}$}
\author{Che-Ming Li$^{1,2,4}$}
\email{cmli@mail.ncku.edu.tw}
\affiliation{$^{1}$Department of Engineering Science, National Cheng Kung University, Tainan 701, Taiwan}
\affiliation{$^{2}$Center for Quantum Frontiers of Research $\&$ Technology, National Cheng Kung University, Tainan 701, Taiwan}
\affiliation{$^{3}$School of Physics, Shandong University, Jinan 250100, China}
\affiliation{$^{4}$Center for Quantum Technology, Hsinchu 30013, Taiwan}

%\affil[*]{cmli@mail.ncku.edu.tw}

\begin{abstract}
Physical processes in the quantum regime possess non-classical properties of quantum mechanics. However, methods for quantitatively identifying such processes are still lacking. Accordingly, in this study, we develop a framework for characterizing and quantifying the ability of processes to cause quantum-mechanical effects on physical systems. We start by introducing a new concept, referred to as quantum process capability, to evaluate the effects of an experimental process upon a prescribed quantum specification. Various methods are then introduced for measuring such a capability. It is shown that the methods are adapted to quantum process tomography for implementation of process capability measure and applicable to all physical processes that can be described using the general theory of quantum operations. The utility of the proposed framework is demonstrated through several examples, including processes of entanglement, coherence, and superposition.
The formalism proposed in this study provides a generic approach for the identification of dynamical processes in quantum mechanics and facilitates the general classification of quantum-information processing.
\end{abstract}

%\begin{document}
%\captionsetup[figure]{labelfont={bf},labelformat={default},labelsep=period,name={Fig.}}
\maketitle

\noindent Physical processes in quantum mechanics attract considerable  interest on account of their unusual characteristics and potential applications. Investigating how and why these processes cannot be explained using classical physics provides an important  insight into the fundamentals of quantum mechanics \cite{Feynman61,Peres95,Breuer&Petruccione02}. By definition, engineering-oriented procedures are physical processes. Then such purely physical investigation inspires the question as to how quantum-mechanical effects can be harnessed to perform practical tasks \cite{Milburn96}. Moreover, feasible techniques for fully exploring the possibilities and limitations of these tasks based on quantum mechanics are still lacking \cite{Feynman82}. The effort to address these issues has revolutionized the conventional methods for engineering physical systems and has greatly advanced the development of quantum technology \cite{Dowling03,O'Brien09,Georgescu12}.

Quantum information processing \cite{Nielsen00} provides a new paradigm for the emerging generation of quantum technologies, such as quantum computation \cite{Ladd10} and quantum communication \cite{Gisin07}. The underlying manipulations of quantum systems are all derived from dynamical processes in quantum mechanics, and range from gate operations \cite{Ladd10,Barenco95} to information storage and protection against the effects of noise \cite{Brennen15}, from the creation of entanglement to teleportation \cite{Bennett93,Xia18} and entanglement swapping \cite{Zukowski93}. Identifying the elementary classes of quantum dynamical processes, therefore, is not only significant in its own right, but is also a fundamental goal in uniting work on quantum information theory \cite{Nielsen00}.

Considerable progress has been made in understanding quantum dynamical processes; in particular, in identifying the quantum properties of the output states
\cite{Lambert10,Li12,Emary13,Brunner14,Baumgratz14,Napoli16,Winter16,Theurer17}.
However, a fully comprehensive analogue for dynamical processes has yet to be found. Moreover, despite the success of theoretical methods in describing the dynamics of quantum systems, such as the quantum operations formalism \cite{Breuer&Petruccione02}, the problem of characterizing the prescribed quantum-mechanical features of dynamical processes in a quantitatively precise manner has yet to be resolved. As a result, it is presently intractable to quantitatively distinguish the different processes in quantum mechanics.

Motivated by this problem, and driven by the desire to ultimately identify all quantum dynamical processes, we present herein a method for quantifying the extent to which a process quantum mechanically behaves and affects a system. We commence by introducing a new concept referred to as the quantum process capability for evaluating the prescribed quantum-mechanical ability of a process. We then introduce two capability measures and a task-oriented capability criterion, which demonstrate that such process evaluation can be quantitatively determined through experimentally feasible means. We show that with the proposed tools, it is possible, for the first time, to quantitatively identify several fundamental types of dynamical process, including processes of entanglement, coherence and superposition.

We begin by systematically characterizing the physical process acting on a system using the quantum operations formalism. In doing so, we assume that the system of interest and its environment are initially in a product state. Furthermore, it is supposed that, after the physical process, the density matrices of the system states, $\rho_{\text{out}}$, can be determined via the state tomography \cite{Vogel89,Leonhardt97}. These experimentally measurable quantities, conditioned on different initial system states, $\rho_{\text{in}}$, are further used in a process tomography algorithm \cite{Nielsen00,Chuang97}. In such a way, the physical process acting on the system can be fully described by a positive Hermitian matrix, referred to hereafter as the process matrix, $\chi_{\text{expt}}$.

The manner in which a system evolves from an arbitrary input state, $\rho_{\text{in}}$, to some process output state, $\rho_{\text{out}}$, is specified by the process matrix $\chi_{\text{expt}}$ through the mapping $\chi_{\text{expt}}(\rho_{\text{in}})=\rho_{\text{out}}$, where this mapping preserves the Hermiticity, trace and positivity of the original  density matrix of the system. With this representation of a dynamical process under our belt, we now turn our attention to defining the concept of quantum process capability.

\section*{Results}

\noindent\textbf{Quantum process capability.} When a process has the ability to show the quantum-mechanical effect on a system prescribed by the specification (for example, entanglement generation), the process is defined as \textit{capable}, and is denoted as $\chi_{\mathcal{C}}$. By contrast, if the process is unable to meet the specification at all,  or is either fully describable using the theory of classical physics or lacks any ability to make the system states quantum mechanical, it is said to be \textit{incapable}.

An incapable process is defined as an operation, $\chi_{\mathcal{I}}$, with the following properties:\\
\noindent (P1) If a process is composed of two cascaded incapable processes: $\chi=\chi_{\mathcal{I}1}\circ\chi_{\mathcal{I}2}$, then process $\chi$ is also incapable, where $\circ$ denotes the concatenation operator.

\noindent (P2) If a process is a linear combination of incapable processes: $\chi$=$\sum_{n} p_{n}\chi_{\mathcal{I}n}$, where $\sum_{n} p_{n}=1$, then $\chi$ is also an incapable process.

\noindent These properties imply that manipulating incapable processes inevitably results in another incapable process.

Furthermore, if the process $\chi_{\text{expt}}$ cannot be described by any incapable processes $\chi_{\mathcal{I}}$ in any way at all, then $\chi_{\text{expt}}$ must be capable regarding the corresponding quantum-mechanical specification.

To place the basic definitions of capable and incapable processes into a wider context, we now introduce a measurable property of a process called the \textit{quantum process capability}, which provides a quantitative understanding of how well  $\chi_{\text{expt}}$ might work, and helps identify processes with prescribed quantum process abilities. The proposed property has applications not only for exploiting the quantum effects behind an unknown process, but also for assisting in the evaluation and improvement of primitive operations for task-oriented purposes. (Note that more general features will be described in the later discussion section.)

The quantum process capability of a process can be quantitatively evaluated using different tools according to the type of specification or subsequently used experimental process. The following discussions propose two methods for evaluating the quantum process capability, namely the capability measures and the task-oriented capability criterion.\\

\noindent\textbf{Capability measure.} We desire to have a tool that can faithfully reflect the features of capable and incapable processes, respectively, and reliably show how the quantum process capability of $\chi_{\text{expt}}$ changes after applying additional operations to the system. Let the capability measure be defined as a function of the process matrix, $C(\chi)$, which has the following three properties:\\
\noindent (MP1) $C(\chi)=0$ if and only if $\chi$ is incapable.

\noindent (MP2) Since incapable processes consist of only incapable ingredients by definition [see (P1) and (P2)], one cannot increase the capability of a process by incorporating additional incapable ones. In other words, it follows that

(2a) The capability measure of $\chi$ monotonically decreases with an incorporated incapable process, i.e., $C(\chi\circ\chi_{\mathcal{I}})\leq C(\chi)$.

(2b) The capability measure reflects the non-increasing capability of a process under stochastic incapable operations: $\sum_{n}p_{n}C(\chi\circ\chi_{\mathcal{I}n})\leq C(\chi)$.

\noindent (MP3) The capability measure is convex, meaning that $C(\sum_{n}p_{n}\chi\circ\chi_{\mathcal{I}n})\leq\sum_{n}p_{n}C(\chi\circ\chi_{\mathcal{I}n})$.

To show how the quantum process capability can be concretely quantified, we now introduce two different types of capability measure, which both satisfy properties (MP1)-(MP3) defined above {(see Methods section).

\noindent{M1. Capability composition $\alpha$}. A process matrix, $\chi_{\text{expt}}$, can be represented as a linear combination of capable and incapable processes, i.e.,
\begin{equation}
\chi_{\text{expt}}=a\chi_{\mathcal{C}}+(1-a)\chi_{\mathcal{I}},\label{com}
\end{equation}
where $a\geq0$. The capability composition of $\chi_{\text{expt}}$ is then defined as
\begin{equation}
\alpha\equiv\min_{\chi_{\mathcal{I}}}a,\label{COM}
\end{equation}
which specifies the minimum amount of capable process that can be found in the experimental process.

In the practical examples presented later in this study, $\alpha$ is obtained by minimizing the following quantity via semi-definite programming (SDP) with MATLAB \cite{Lofberg,sdpsolver}: $\alpha=\min_{\tilde{\chi}_{\mathcal{I}}}[1-\text{tr}(\tilde{\chi}_{\mathcal{I}})]$. Note that the solution is obtained under a set of specified conditions for the incapable process, $D(\tilde{\chi}_{\mathcal{I}})$, such that $\chi_{\text{expt}}-\tilde{\chi}_{\mathcal{I}}=\tilde{\chi}_{\mathcal{C}}\geq0$. Here, $\tilde{\chi}_{\mathcal{I}}$ and $\tilde{\chi}_{\mathcal{C}}$ are both unnormalized process matrices with $\text{tr}(\tilde{\chi}_{\mathcal{I}})=\text{tr}((1-a)\chi_{\mathcal{I}})=1-a$ and $\text{tr}(\tilde{\chi}_{\mathcal{C}})=\text{tr}(a\chi_{\mathcal{C}})=a$, respectively. Rephrasing the process matrices in terms of unnormalized matrices reduces the number of variables to make the problem solvable using SDP. The number of variables is determined by the number of elements in a process matrix. For example, there are at least 16 variables that need to be solved in a single qubit system. For different capabilities, it may need more variables to describe the constraints $D(\tilde{\chi}_{\mathcal{I}})$. See Methods.
$D(\tilde{\chi}_{\mathcal{I}})$
places constraints on the process matrix construction of the incapable process in the process tomography algorithm, which specifies how the input and output states for the process tomography behave under an incapable process.

\noindent{M2. Capability robustness $\beta$}. An experimental process, $\chi_{\text{expt}}$, can become incapable by mixing with noise, i.e.,
\begin{equation}
\frac{\chi_{\text{expt}}+b\chi'}{1+b}=\chi_{\mathcal{I}},\label{rob}
\end{equation}
where $b\geq 0$ and $\chi'$ is the noise process. The capability robustness of $\chi_{\text{expt}}$ can be defined as the minimum amount of noise which must be added such that $\chi_{\text{expt}}$ becomes $\chi_{\mathcal{I}}$, i.e.,
\begin{equation}
\beta\equiv\min_{\chi'}b.\label{ROB}
\end{equation}
In practical cases, $\beta$ can be obtained by using SDP to solve $\beta=\min_{\tilde{\chi}_{\mathcal{I}}}[\text{tr}(\tilde{\chi}_{\mathcal{I}})-1]$, under $D(\tilde{\chi}_{\mathcal{I}})$ such that $\text{tr}(\tilde{\chi}_{\mathcal{I}})\geq1,\ \tilde{\chi}_{\mathcal{I}}-\chi_{\text{expt}}\geq0$, which ensure that $\beta\geq0$ and $\chi'$ is positive semi-definite, respectively.\\

\noindent\textbf{Quantum criterion.} When $\chi_{\text{expt}}$ is created with respect to a target process, $\chi_{\text{target}}$, that is defined as capable for engineering-oriented purposes \cite{Nielsen00,Ladd10,Gisin07}, the process fidelity, $F_{\text{expt}}\equiv \text{tr}(\chi_{\text{expt}}\chi_{\text{target}})$, enables us to examine the similarity between them. In particular, $\chi_{\text{expt}}$ is judged to have a capability close to that of the target process if it goes beyond the best mimicry by incapable processes to $\chi_{\text{target}}$, i.e.,
\begin{equation}
F_{\text{expt}}>F_{\mathcal{I}}\equiv\max_{\chi_{\mathcal{I}}}[\text{tr}(\chi_{\mathcal{I}}\chi_{\text{target}})],\label{F}
\end{equation}
meaning that $\chi_{\text{expt}}$ is a faithful operation which cannot be simulated by any incapable processes. Note that $F_{\mathcal{I}}$ in (\ref{F}) can be evaluated by performing the following maximization task with SDP: $F_{\mathcal{I}}=\max_{\tilde{\chi}_{\mathcal{I}}}[\text{tr}(\tilde{\chi}_{\mathcal{I}}\chi_{\text{target}})]$, under $D(\tilde{\chi}_{\mathcal{I}})$ such that $\text{tr}(\tilde{\chi}_{\mathcal{I}})=1$. Moreover, from a reasonable engineering-oriented perspective, we consider $\chi_{\text{target}}$ capable and to be the default, i.e., $F_{\mathcal{I}}<1$. Note that the criterion (\ref{F}) does not satisfy (MP1)-(MP3). If the process fidelity under the criterion (\ref{F}) is used to serve as a capability measure, say $C(\chi_{\text{expt}})\equiv F_{\text{expt}}-F_{\mathcal{I}}$, this fidelity measure will not be necessarily equal to zero when the $\chi_{\text{expt}}$ is an incapable process. The reason is that $F_{\mathcal{I}}$ is the optimal value for all incapable processes, which means that the fidelity criterion (\ref{F}) does not satisfy (MP1). Since (MP2) and (MP3) are derived from (MP1), such criterion also does not satisfies (MP2) and (MP3).\\

\noindent\textbf{Capability examples.}
Let the following capable processes be used to demonstrate the formalism described above:

\noindent E1. Non-classical dynamics. As defined in ref. ~\cite{Hsieh17}, a non-classical dynamics is a dynamical process that cannot be explained using a classical picture: the initial system is considered a physical object with properties satisfying the assumption of classical realism, and the system thus evolves according to classical stochastic theory.

Non-classical dynamics can be quantified and used as the requirement for the non-classical manipulation of a system \cite{Hsieh17}, e.g., the fusion of entangled photon pairs \cite{Pan12}. In the present context of quantum process capability, non-classical dynamics is capable of going beyond the generic model of classical dynamics, compared to incapable classical processes. The quantifications of non-classical processes introduced in ref. ~\cite{Hsieh17} reveal the capability of non-classical dynamics, and include the capability composition $\alpha$ (\ref{COM}), the capability robustness $\beta$ (\ref{ROB}) and the capability criterion (\ref{F}).

\noindent E2. Entanglement generation. Creating entanglement is a crucial dynamical process in both quantum mechanics and quantum-information processing \cite{Nielsen00,Ladd10}. However, the means to quantify the ability of a process to generate the entanglement of two qubits remains unclear.

Let entanglement creation be defined as a capable process, $\chi_{\text{ent}}$. Any process that merely preserves the separability of a quantum system is therefore said to be incapable, and is denoted by $\chi_{\mathcal{I},\text{ent}}$. The capability measures of $\chi_{\text{ent}}$, such as $\alpha$ and $\beta$, confirm that $C(\chi_{\text{ent}})>0$. By contrast, the capability measure of $\chi_{\mathcal{I},\text{ent}}$ is minimum, i.e., $C(\chi_{\mathcal{I},\text{ent}})=0$. The concrete set of constraints acting on the incapable process $D(\tilde{\chi}_{\mathcal{I},\text{ent}})$ in calculating $\alpha$, $\beta$ and $F_{\mathcal{I}}$ is given by
\begin{equation}
\tilde{\chi}_{\mathcal{I},\text{ent}}(\rho_{\text{in}})\geq0\ \forall \rho_{\text{in}};(\tilde{\chi}_{\mathcal{I},\text{ent}}(\rho_{\text{in}}))^{\text{PT}}\geq0\ \forall \rho_{\text{in}}\in s_{\text{sep}},\label{Dent}
\end{equation}
where $s_{\text{sep}}$ denotes the set of separable states. The first  constraint in equation~(\ref{Dent}) ensures that the output states are positive semi-definite for all the input states required in the process tomography algorithm. As will be illustrated below, this condition is necessary for all incapable processes. The second constraint is based on the positive partial transpose (PPT) criterion \cite{Peres96,Horodecki96}, and guarantees that if the input states are separable states, the output states are separable states as well. In other words, the PPT criterion stipulates that an output state $\rho_{\text{out}}$ after partial transposition (PT) is positive semi-definite, i.e., $\rho_{\text{out}}^{\text{PT}}\geq 0$, if and only if the state $\rho_{\text{out}}$ is separable.

\noindent E3. Coherence creation and preservation. Quantum coherence is one of the main features of quantum systems \cite{Li12,Baumgratz14,Napoli16,Winter16}, and is the main power behind quantum technology \cite{Dowling03,O'Brien09}. If the density matrix of a $d$-dimensional system is not diagonal in a given orthonormal basis $\{\ket{j}|j=0,1,...,d-1\}$, then the system is said to possess coherence with respect to basis $\{\ket{j}\}$ \cite{Baumgratz14}. A mixture consisting  only of the basis states of the form: $\sum_{j=0}^{d-1}p_{j}\ket{j}\!\bra{j}$, is then said to be an incoherent state.

Coherence creation and preservation are essential in performing state preparation and manipulation in quantum engineering \cite{Nielsen00,Ladd10,Gisin07,Brennen15}. They represent two different abilities of capable processes, and are denoted hereafter as $\chi_{\text{cre}}$ and $\chi_{\text{pre}}$, respectively. The capability measure of coherence creation shows that $C(\chi_{\text{cre}})>0$. Compared with $\chi_{\text{cre}}$, an incapable process, $\chi_{\mathcal{I},\text{cre}}$, cannot create coherence, i.e., incoherent states remain incoherent, and hence the coherence capability measure  is minimum, $C(\chi_{\mathcal{I},\text{cre}})=0$. To evaluate $\alpha$, $\beta$ and $F_{\mathcal{I}}$, for  a process incapable of coherence creation, $D(\tilde{\chi}_{\mathcal{I},\text{cre}})$ is set as
\begin{equation}
\tilde{\chi}_{\mathcal{I},\text{cre}}(\rho_{\text{in}})\geq0\ \forall \rho_{\text{in}};\tilde{\chi}_{\mathcal{I},\text{cre}}(\rho_{\text{in}})\in \tilde{s}_{\text{incoh}}\ \forall \rho_{\text{in}}\in s_{\text{incoh}},
\end{equation}
where $\tilde{s}_{\text{incoh}}$ and $s_{\text{incoh}}$ denote the sets of unnormalized incoherent states and incoherent states regarding the basis $\{\ket{j}\}$, respectively. In other words, if the input states are incoherent states, then the output states must also be incoherent states.

Regarding the capability of coherence preservation, {$C(\chi_{\text{pre}})$ has a positive value for a capable process. By contrast, an incapable process causes the states to decohere, i.e., $C(\chi_{\mathcal{I},\text{pre}})=0$. The constrain set, $D(\tilde{\chi}_{\mathcal{I},\text{pre}})$, for an incapable process is formulated as
\begin{equation}
\tilde{\chi}_{\mathcal{I},\text{pre}}(\rho_{\text{in}})\geq0,\  \tilde{\chi}_{\mathcal{I},\text{pre}}(\rho_{\text{in}})\in \tilde{s}_{\text{incoh}}\ \forall \rho_{\text{in}}.
\end{equation}
In other words, all of the output states are incoherent states, regardless of the input state $\rho_{\text{in}}$. It is noted that this is different from the coherence creation case, in which the input states are restricted to incoherent states.

\noindent E4. Superposition of quantum states. The superposition principle of quantum mechanics \cite{Theurer17,Dirac30} describes how any two or more quantum states can be superposed together to form another valid quantum state (and \textit{vice versa}). The superposition of quantum states therefore generalizes the capability of coherence creation in the orthonormal basis $\{\ket{j}\}$ to the ability to superpose states in a normalized and linear independent basis $\{\ket{h_{j}}|j=0,1,...,d-1\}$. Superposition states cannot be explained by any mixture of the basis states: $\sum_{j=0}^{d-1}p_{j}\ket{h_{j}}\!\bra{h_{j}}$, and are hence referred to as} superposition free states \cite{Theurer17}.

For a capable superposition process, the capability measure has a positive value, $C(\chi_{\text{sup}})>0$. By contrast, for an incapable process, $\chi_{\mathcal{I},\text{sup}}$, i.e., a process which merely preserves the mixture in the superposition free states, the capability measure is given as
 $C(\chi_{\mathcal{I},\text{sup}})=0$. The constraints on the incapable process, $D(\tilde{\chi}_{\mathcal{I},\text{sup}})$ are stated as
\begin{equation}
\tilde{\chi}_{\mathcal{I},\text{sup}}(\rho_{\text{in}})\geq0\  \forall \rho_{\text{in}};\tilde{\chi}_{\mathcal{I},\text{sup}}(\rho_{\text{in}})\!\in\!\tilde{s}_{\text{supf}}\ \forall \rho_{\text{in}}\!\in\! s_{\text{supf}},
\end{equation}
where $\tilde{s}_{\text{supf}}$ and $s_{\text{supf}}$ denote the sets of unnormalized superposition free states and superposition free states under the basis $\{\ket{h_{j}}\}$, respectively.  $D(\tilde{\chi}_{\mathcal{I},\text{sup}})$ indicates that if $\rho_{\text{in}}$ is a superposition free state, then the output state must also be superposition free.\\

\noindent\textbf{Demonstration of the process capability measures.}
We now provide an explicit example illustrating how a dynamical process of interest can be identified upon the prescribed quantum specification (i.e., E1-E4). We consider the case of a composite system consisting of two qubits. Let the qubits be coupled via an interaction, equivalent to the quantum Ising model, of the Hamiltonian \cite{Briegel01}: $H_{\text{int}}=1/2\sum_{j,k=0}^1(-1)^{jk}\ket{jk}\!\bra{jk}$, which is an important primitive for creating cluster states in a one-way quantum computer \cite{Raussendorf01}. The composite two-qubit evolution can be represented by an unitary transform $U(t)=\sum_{j,k=0}^1e^{0.5(-1)^{jk}it}\ket{jk}\!\bra{jk}$. When $t=\pi$, the unitary transform $U(\pi)=\sum_{j,k=0}^1(-1)^{jk}\ket{jk}\!\bra{jk}$ can be considered as an implementation of the controlled-$Z$ gate. Moreover, assume that one of the qubits is depolarized at a rate $\gamma$. The depolarization of a qubit can be represented in the form $\rho_{\text{final}}=e^{-\gamma t}\rho_{\text{initial}}+(1-e^{-\gamma t})\hat{I}/2$, where $\hat{I}$ is identity matrix. By using the process tomography algorithm, the process matrix of such depolarization is
\begin{equation}
\chi_{\text{dep}}= \frac{1}{2}\left[ \begin{matrix}
    \frac{1+e^{-\gamma t}}{2} & 0 & 0 & e^{-\gamma t}\\
    0 & \frac{1-e^{-\gamma t}}{2} & 0 & 0\\
    0 & 0 & \frac{1-e^{-\gamma t}}{2} & 0\\
    e^{-\gamma t} & 0 & 0 & \frac{1+e^{-\gamma t}}{2}
    \end{matrix}
\right].\nonumber
\end{equation}
Given $\gamma$=0.02, Figs.~\ref{CZtwo}a and b illustrate the variations of $\alpha$ and $\beta$,  respectively, for the five different capabilities described above with the qubit interaction time, $t$. The set of specified conditions for different capabilities for the incapable process, $D(\tilde{\chi}_{\mathcal{I}})$, used to evaluate $\alpha$, $\beta$, and $F_{\mathcal{I}}$ for this example are given in Methods section.
Figure~\ref{CZtwo}c compares the process fidelity, $F_{\text{expt}}$, with the four different thresholds $F_{\mathcal{I}}$ of the capable processes, with respect to a target process, the controlled-$Z$ gate. When $F_{\text{expt}}=1$, it means the $\chi_{\text{expt}}(t)$ arrives at the target process, the controlled-$Z$ gate. There are oscillations of $\alpha$, $\beta$, and $F_{\text{expt}}$ in Fig.~\ref{CZtwo}. The oscillation period is 2$\pi$ for $U(t)=U(t+2\pi)$, which is determined by the Hamiltonian. If we double the values of the interaction intensity in $H_{\text{int}}$ and the Hamiltonian become $H'_{\text{int}}=\sum_{j,k=0}^1(-1)^{jk}\ket{jk}\!\bra{jk}$, the oscillation period will become $\pi$ in our example. An observation of Fig.~\ref{CZtwo} yields several important conclusions regarding process identification and classification, as described in the following.

First, compared to $\alpha$, the capability robustness measure $\beta$ provides a clearer distinction between the various processes under different interaction periods. This distinction can be realized by characterizing \textit{how close} $\chi_{\text{expt}}(t)$ \textit{is to an incapable process} in the sense of how large the minimum amount of noise $\beta$ is required to be to make $\chi_{\text{expt}}(t)$ incapable (see Methods). %\cite{alphaDistinction}.

Second, while the process considered in Fig.~\ref{CZtwo} has no ability to create coherence, it enables the two qubits to be entangled. Such a result implies that the coherence of the output entangled qubits is not generated by the process $\chi_{\text{expt}}(t)$ itself, but is transformed from the input states. In other words, $\chi_{\text{expt}}(t)$ has the ability to convert the coherence of the input states into an entanglement of the qubits at specific qubit interaction times. The efficiency of this coherence conversion process can be determined by the capability of entanglement generation, as evaluated by $\beta$, for example (see Methods).

Third, superposition can be performed either by purely unitary evolution or by quantum jumps of depolarization. The latter phenomenon can be seen in Fig.~\ref{CZtwo}b, in which $\chi_{\text{expt}}(t)$ for large $t$ (where the depolarization dominates the dynamics) possesses this ability, as described by $\beta>0$.

Finally, since superposition can be implemented using incoherent operations, the capability of superposition does not show how efficiently the coherence of the input states $\sum_{j=0}^{d-1}p_{j}\ket{h_{j}}\!\bra{h_{j}}$ can be converted into a superposition of the basis states $\ket{h_{j}}$. (Note that this fact can be understood in an analogous fashion to that used to understand the conversion of the coherence of the input states into entanglement) (see Methods).

\begin{figure}
\includegraphics[width=8.5cm]{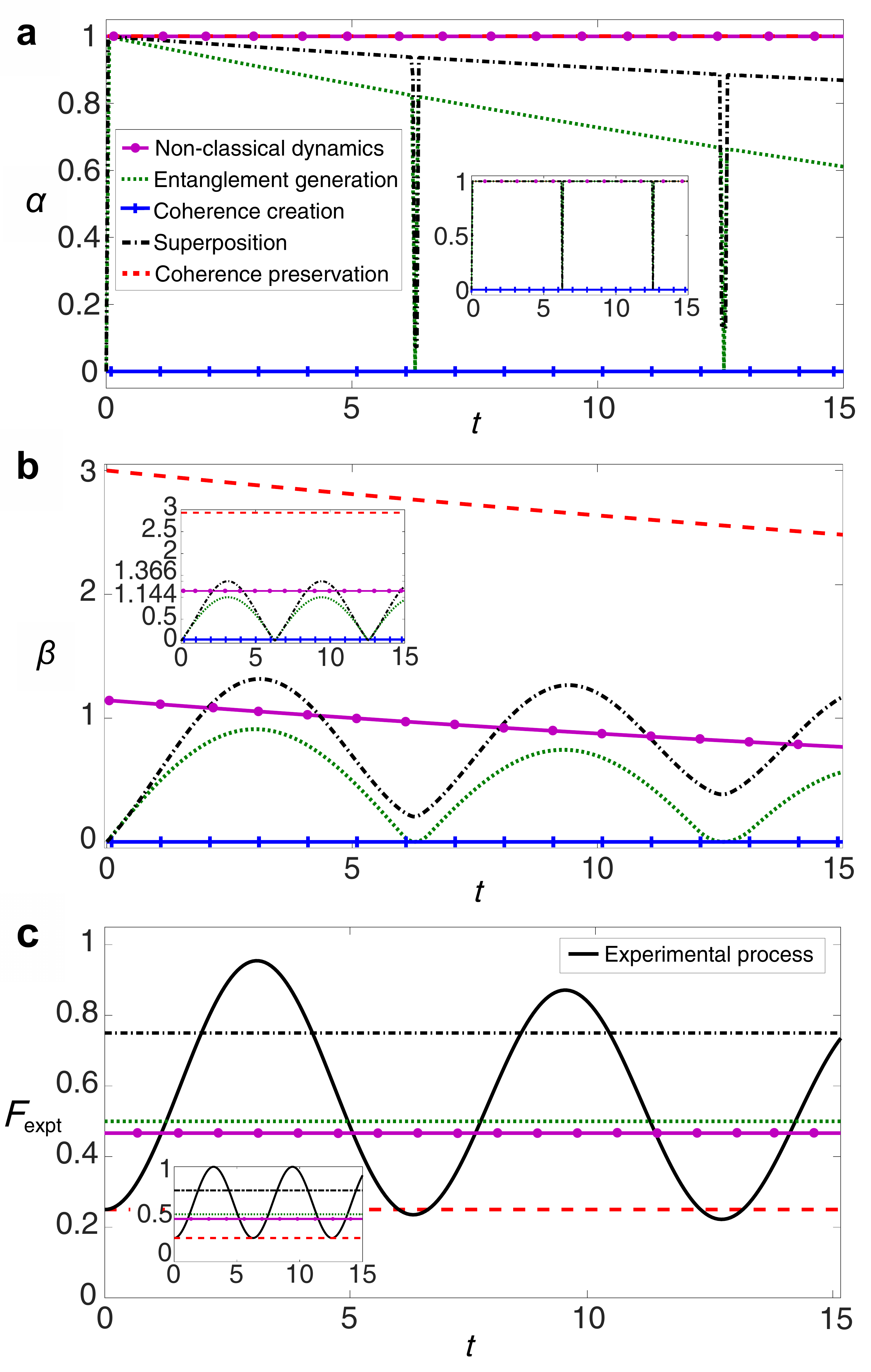}
\caption{Evaluating dynamical processes with process capability measures. The five different capabilities in the dynamics $\chi_{\text{expt}}(t)$ of two coupled qubits under a single-qubit depolarizing channel are examined using the capability measures (\textbf{a}) $\alpha$, (\textbf{b}) $\beta$, and (\textbf{c}) capability criterion. Note that the coherence and superposition of the states are defined in the bases $\{\ket{00},\ket{01},\ket{10},\ket{11}\}$ and $\{\ket{00},(\ket{00}+\ket{01})/\sqrt{2},(\ket{00}+\ket{10})/\sqrt{2},(\ket{00}+\ket{11})/\sqrt{2}\}$, respectively. For comparison with the capability change over time under qubit depolarization, the insets show the corresponding cases without noise. The depolarizing rate $\gamma$ affects the curves of $\alpha$, $\beta$, and $F_{\text{expt}}$. Here we set $\gamma=0.02$ in the present example. It is worth noting that $\chi_{\text{expt}}(t)$ can serve as a controlled-$Z$ (CZ) gate at a proper interaction time \cite{Raussendorf01}. When setting this gate operation as the target process, the process fidelity $F_{\text{expt}}$ varies with time. As indicated in (\textbf{c}), the capability thresholds $F_{\mathcal{I}}$ for superposition, entanglement generation, non-classical dynamics, and coherence preservation are: $0.750$, $0.500$, $0.467$ and $0.250$, respectively. Since the CZ gate can be described by incapable processes $\chi_{\mathcal{I},\text{cre}}$, and is not a proper target process for coherence creation,  such process capability is not considered in (\textbf{c}). Note that, in (\textbf{a}), for coherence preservation, the process is always a capable process, since the depolarization only acts on one of the qubits and another qubit can still preserve coherence of single qubit. According to the definition of the incapable process of coherence preservation (i.e., \textit{all the output states} must be incoherent states), this process is always capable process of coherence preservation.}
\label{CZtwo}
\end{figure}

\section*{Discussion}

\noindent The quantum process capability framework introduced above possesses several important features for practical utility and extensions, as described in the following.

\noindent (D1) Since constructing the process matrix is experimentally feasible, the process capability can be readily quantified in various present experiments. The quantum operations formalism underlying the process matrix is a general tool for describing the dynamics experienced by closed or open quantum systems in a wide variety of physical scenarios. Our formalism is therefore applicable to all physical processes described by the general theory of quantum operations, including, but not limited to, the fundamental processes postulated in quantum mechanics \cite{Peres95,Nielsen00}, the dynamics of energy transfer systems \cite{Yuen-Zhou11},  the task-orientated processes associated with quantum information \cite{Ladd10,Gisin07,Pan12}, and the atomic-physics and quantum-optics experiments on a chip \cite{You11}.

\noindent (D2) In addition to the capable processes illustrated above (i.e., E1-E4), the present framework can also be used to explore other types of quantum process capability, including the creation of genuine multipartite entanglement \cite{Guhne09}, Einstein-Podolsky-Rosen steering \cite{Wiseman07}, Bell non-locality \cite{Brunner14}, and genuine multipartite quantum steering \cite{He13,Li15}.

\noindent (D3) Process identification and classification. As distinct process capabilities are given, dynamical processes can be identified and classified accordingly using the capability measures, such as the capability robustness. This is helpful in uniting existing works on quantum information under a given type of quantum process capability, for example, the preservation of quantumness \cite{Hsieh17} or the coherence of quantum information \cite{Nielsen00,Ladd10,Gisin07,Brennen15}.

\noindent (D4) Benchmark for process engineering. The capability criterion (\ref{F}) sets a level $F_{\mathcal{I}}$ of quantum-mechanical quality that can be used as a standard when comparing other experimental processes for task-oriented purposes \cite{Nielsen00,Ladd10,Gisin07}; as illustrated in Fig.~\ref{CZtwo}c.

\noindent (D5) Insightful description of processes. A given quantum process capability may  be linked in some way with other concepts. For example, the capability of non-classical dynamics can be used to reveal the non-Markovianity of dynamical processes \cite{Hsieh17,Wang18,Vega17}. Alternatively, the capability of entanglement generation also provides an understanding of the efficiency of coherence conversion. The framework proposed herein thus facilitates a more comprehensive understanding of the characteristics of quantum dynamics.

\noindent (D6) Maximum extraction of resources from processes. The quantum effects of a process on the system inputs, $\rho_{\text{in}}$, as revealed by the output states, $\rho_{\text{out}}$, are usually considered as quantum resources, such as entanglement \cite{Vedral98,Plenio07} and coherence \cite{Baumgratz14,Napoli16}. The maximum amount of resources that can be extracted from a given process, $\chi_{\text{expt}}$, can be quantified in terms of the quantum process capability, $C(\chi_{\text{expt}})$, using an appropriate capability measure.  The reason for this lies in the fact that the quantum process capability provides an optimal description of the prescribed quantum specification in $\chi_{\text{expt}}$ in the sense that the framework proposed in this study does not depend on any specifics of the states being processed (see Methods section).

In conclusion, we have developed a novel formalism for performing the quantitative identification of quantum dynamical processes. The concept of quantum process capability and its various quantifications have been introduced to evaluate the prescribed quantum-mechanical features of a dynamical process. The ability to quantify the ability of a process in terms of its quantum process capabilities makes it possible to discriminate quantitatively between different dynamical processes. Overall, the rigorous framework of quantum process capability proposed in this study provides the means to go beyond the usual analysis of state characteristics and to approach the goal of uniting work on quantum information theory.

Future studies will aim to improve the performance and scalability of the process tomography \cite{Shabani11} underlying the proposed formalism, including scenarios such as where the measurement outcomes are continuous and unbound, e.g., as for nanomechanical resonators \cite{Johansson14}. It is anticipated that the enhanced framework will thus facilitate the novel recognition and classification of physical processes with quantum process capability.\\

\section*{Methods}

%We first show that the capability composition measure $\alpha$ and capability robustness measure $\beta$ possess the three required properties of capability measures, namely, (MP1)-(MP3). The difference between $\alpha$ and $\beta$ is discussed as well. We then present the sets of specified conditions, $D(\tilde{\chi}_{\mathcal{I}})$, for the incapable processes $\tilde{\chi}_{\mathcal{I}}$ illustrated in the capability examples, E1-E4, and their corresponding explicit illustrations in Fig.~\ref{CZtwo}. We also describe the ability of the dynamical process in Fig.~\ref{CZtwo} to convert the coherence of the input states into an entanglement and superposition of the qubits. Why the capability of entanglement generation can be used to examine the coherence conversion efficiency is shown. Finally, we prove that the maximum amount of resources that can be extracted from a given process can be quantified in terms of the quantum process capability.\\

\noindent\textbf{Basic properties of capability composition $\alpha$}. The capability composition measure $\alpha$ possess the three required properties of capability measures, namely, (MP1)-(MP3).

\noindent \text{(MP1)}:
From the definition of capability composition, equations~(\ref{com}) and (\ref{COM}), it follows directly that (MP1) is a fulfilled property.

\noindent \text{(MP2)}: (2b)

Property (2a) can be proven by properties (2b) and (MP3). We start by proving (2b).

We consider a process consisting of the experimental process \cite{Chuang97,Nielsen00}, $\chi_{\text{expt}}$, and an incapable process, $\chi_{\mathcal{I}}$, of the form
\begin{equation}
\chi_{\text{expt}}\circ\chi_{\mathcal{I}}=\sum_{n} p_{n}\chi_{\text{expt}}\circ\chi_{\mathcal{I}n},\label{ten}
\end{equation}
where each $\chi_{\mathcal{I}n}$ is an
incapable process and $\{p_{n}\}$ is the corresponding probability distribution with $\sum_{n}p_{n}=1$. Using equations~(\ref{com}) and (\ref{COM}), the process $\chi_{\text{expt}}\circ\chi_{\mathcal{I}n}$ can be represented as
\begin{equation}
\chi_{\text{expt}}\circ\chi_{\mathcal{I}n}=\alpha\chi_{\mathcal{C}}\circ\chi_{\mathcal{I}n}+(1-\alpha)\chi_{\mathcal{I}}\circ\chi_{\mathcal{I}n}.
\end{equation}
Note that, while $\alpha$ is the capability composition of $\chi_{\text{expt}}$, i.e., the optimal result for $\chi_{\text{expt}}$, it is not necessarily the optimal result for $\chi_{\text{expt}}\circ\chi_{\mathcal{I}n}$. Since $\chi_{\mathcal{I}n}$ and $\chi_{\mathcal{I}}$ are both incapable processes, $\chi_{\mathcal{I}}\circ\chi_{\mathcal{I}n}$ must also be an incapable process.

For $\chi_{\mathcal{C}}\circ\chi_{\mathcal{I}n}$, there exists a possibility that the process is not capable, i.e., it contains a certain non-vanished amount of incapable process. Therefore, the minimum amount of capable process that can be found in $\chi_{\text{expt}}\circ\chi_{\mathcal{I}n}$ cannot be greater than $\alpha$. In other words, it follows that it is possible to find specific capable and incapable processes, denoted by $\chi_{\mathcal{C}n'}$ and $\chi_{\mathcal{I}n'}$, respectively, which yield the following decomposition:
\begin{equation}
\chi_{\text{expt}}\circ\chi_{\mathcal{I}n}=\alpha_{n}\chi_{\mathcal{C}n'}+(1-\alpha_{n})\chi_{\mathcal{I}n'}.
\end{equation}
Since $\alpha_{n}$ is the capability composition of $\chi_{\text{expt}}\circ\chi_{\mathcal{I}n}$, i.e., it is optimal for $\chi_{\text{expt}}\circ\chi_{\mathcal{I}n}$, it follows that
\begin{equation}
\alpha_{n}\leq \alpha,
\end{equation}
which implies that
\begin{equation}
\sum_{n}p_{n}\alpha_{n}\leq\alpha.
\end{equation}
Thus, we conclude that the capability composition measure satisfies the property, $\sum_{n}p_{n}C(\chi_{\text{expt}}\circ\chi_{\mathcal{I}n})\leq C(\chi_{\text{expt}})$.

\noindent (MP3): The process $\chi_{\text{expt}}\circ\chi_{\mathcal{I}}$ (\ref{ten}) can be reformulated as
\begin{equation}
\chi_{\text{expt}}\circ\chi_{\mathcal{I}}=a_{\mathcal{I}}\chi_{\mathcal{C'}}+(1-a_{\mathcal{I}})\chi_{\mathcal{I'}},
\end{equation}
where $\chi_{\mathcal{C'}}=\sum_{n}p_{n}\alpha_{n}\chi_{\mathcal{C}n'}/a_{\mathcal{I}},$
$\chi_{\mathcal{I'}}=\sum_{n}p_{n}(1-\alpha_{n})\chi_{\mathcal{I}n'}/(1-a_{\mathcal{I}}),$ and
$a_{\mathcal{I}}=\sum_{n}p_{n}\alpha_{n}$.
Note that $a_{\mathcal{I}}$ is not necessarily optimal for $\chi_{\text{expt}}\circ\chi_{\mathcal{I}}$.

Since each process $\chi_{\mathcal{I}n'}$ is an incapable process, $\chi_{\mathcal{I}'}$ is also an incapable process. In accordance with the definition of capability composition, we have $\alpha_{\mathcal{I}}=C(\sum_{n}p_{n}\chi_{\text{expt}}\circ\chi_{\mathcal{I}n})$ and
$\alpha_{\mathcal{I}}\leq a_{\mathcal{I}}$. Moreover, since $\alpha_{n}=C(\chi_{\text{expt}}\circ\chi_{\mathcal{I}n})$, we finally get
\begin{equation}
\alpha_{\mathcal{I}}\leq \sum_{n}p_{n}\alpha_{n},
\end{equation}
which satisfies the property, $C(\sum_{n}p_{n}\chi_{\text{expt}}\circ\chi_{\mathcal{I}n})\leq \sum_{n}p_{n}C(\chi_{\text{expt}}\circ\chi_{\mathcal{I}n})$.

\noindent \text{(MP2)}: (2a)

From (2b) and (MP3), we conclude that
\begin{equation}
\alpha_{\mathcal{I}}\stackrel{(\text{MP3})}{\leq} \sum_{n}p_{n}\alpha_{n}\stackrel{(\text{2b})}\leq \alpha,
\end{equation}
i.e., $C(\chi_{\text{expt}}\circ\chi_{\mathcal{I}})\leq C(\chi_{\text{expt}})$.\\

Here we take the capability composition $\alpha$ of entanglement generation for example to demonstrate the three properties (MP1) - (MP3) for capability measure. Let us consider the instance of coupled qubits discussed in Fig.~\ref{CZtwo}. As shown in the inset of Fig.~\ref{CZtwo}a for time evolution without noise effect, when $t=0$, the process $\chi_{\text{expt}}(0)$ is identity process which has no capability to generate entanglement from separable states. Then $\chi_{\text{expt}}(0)$ is an incapable process and its $C(\chi_{\text{expt}}(0))$ is $0$ (MP1). If we incorporate the incapable process $\chi_{\text{expt}}(0)$ into the capable process $\chi_{\text{expt}}(\pi)$, we find that the incapable process does not increase the capability composition $\alpha$,
\begin{equation}
C(\chi_{\text{expt}}(\pi)\circ\chi_{\text{expt}}(0))= C(\chi_{\text{expt}}(\pi))=1, \nonumber
\end{equation}
which satisfies property (2a) in (MP2). Since other incapable processes $\chi_{\text{expt}}(2n\pi)$ do not increase the capability composition, i.e., $C(\chi_{\text{expt}}(\pi)\circ\chi_{\text{expt}}(2n\pi))=1$, it satisfies property (2b) in (MP2), $\sum_{n}p_{n}C(\chi_{\text{expt}}(\pi)\circ\chi_{\text{expt}}(2n\pi))\leq C(\chi_{\text{expt}}(\pi))$.
To show an example of the property (MP3), we mix $\chi_{\text{expt}}(\pi)\circ\chi_{\text{expt}}(0)$ and $\chi_{\text{expt}}(\pi)\circ\chi_{\text{expt}}(2\pi)$ together, where $\chi_{\text{expt}}(0)$ and $\chi_{\text{expt}}(2\pi)$ are both incapable processes. After mixing, the capability composition does not increase,
\begin{equation}
\begin{split}
&\frac{1}{2}C(\chi_{\text{expt}}(\pi)\circ\chi_{\text{expt}}(0))+\frac{1}{2}C(\chi_{\text{expt}}(\pi)\circ\chi_{\text{expt}}(2\pi))\\ &= C(\frac{1}{2}\chi_{\text{expt}}(\pi)\circ\chi_{\text{expt}}(0)+\frac{1}{2}\chi_{\text{expt}}(\pi)\circ\chi_{\text{expt}}(2\pi))=1, \nonumber
\end{split}
\end{equation}
which satisfies (MP3).

The properties (MP1)-(MP3) also hold for the time evolution under the noise effect. In Fig.~\ref{CZtwo}a, when $t=2\pi$, the process $\chi_{\text{expt}}(2\pi)$ is a process that has no capability to generate entanglement from separable states, so $\chi_{\text{expt}}(2\pi)$ is an incapable process and its $C(\chi_{\text{expt}}(2\pi))$ is $0$ (MP1). The capability composition of a capable process $\chi_{\text{expt}}(\pi)$ does not increase by incorporating additional incapable process $\chi_{\text{expt}}(2\pi)$, i.e.,
\begin{equation}
C(\chi_{\text{expt}}(\pi)\circ\chi_{\text{expt}}(2\pi))=0.742< C(\chi_{\text{expt}}(\pi))=0.909,  \nonumber
\end{equation}
and satisfies property (2a) in (MP2). The example for property (2b) in (MP2) can be known from the case,
\begin{equation}
\begin{split}
&C(\frac{1}{2}\chi_{\text{expt}}(\pi)\circ\chi_{\text{expt}}(2\pi)+\frac{1}{2}\chi_{\text{expt}}(\pi)\circ\chi_{\text{expt}}(4\pi))=0.669\\ &< C(\chi_{\text{expt}}(\pi))=0.909,  \nonumber
\end{split}
\end{equation}
where $\chi_{\text{expt}}(2\pi)$ and $\chi_{\text{expt}}(4\pi)$ are both incapable processes.
If we mix $\chi_{\text{expt}}(\pi)\circ\chi_{\text{expt}}(2\pi)$ and $\chi_{\text{expt}}(\pi)\circ\chi_{\text{expt}}(4\pi)$ together, the mixed process satisfies the property (MP3),
\begin{equation}
\begin{split}
&\frac{1}{2}C(\chi_{\text{expt}}(\pi)\circ\chi_{\text{expt}}(2\pi))+\frac{1}{2}C(\chi_{\text{expt}}(\pi)\circ\chi_{\text{expt}}(4\pi))\\ &= C(\frac{1}{2}\chi_{\text{expt}}(\pi)\circ\chi_{\text{expt}}(2\pi)+\frac{1}{2}\chi_{\text{expt}}(\pi)\circ\chi_{\text{expt}}(4\pi))\\
&=0.669 ,\nonumber
\end{split}
\end{equation}
that the mixing does not increase capability composition.

\noindent\textbf{Basic properties of capability robustness $\beta$.} The capability robustness measure $\beta$ has the three properties of capability measures, (MP1)-(MP3).

\noindent (MP1):
From the definition of capability robustness, equations~(\ref{rob}) and (\ref{ROB}), it is apparent that (MP1) is a fulfilled property.

\noindent (MP2): (2b)

From the definition of capability robustness, equations~(\ref{rob}) and (\ref{ROB}), it is additionally clear that
\begin{equation}
\chi_{\text{expt}}=(1+\beta)\chi_{\mathcal{I}}-\beta\chi'.\label{eighteen}
\end{equation}
From equation~(\ref{ten}), combined with equation~(\ref{eighteen}), we have
\begin{equation}
\chi_{\text{expt}}\circ\chi_{\mathcal{I}n}=(1+\beta)\chi_{\mathcal{I}}\circ\chi_{\mathcal{I}n}-\beta\chi'\circ\chi_{\mathcal{I}n}.
\end{equation}
Since $\chi_{\mathcal{I}n}$ and $\chi_{\mathcal{I}}$ are both incapable processes, $\chi_{\mathcal{I}}\circ\chi_{\mathcal{I}n}$ must also be an incapable process.
It is worth noting that, while $\beta$ is the robustness of $\chi_{\text{expt}}$, i.e., it is an optimal value for $\chi_{\text{expt}}$, it is not necessarily the optimal value for the resulting process, $\chi_{\text{expt}}\circ\chi_{\mathcal{I}n}$.

In general, the quantum process capability cannot be increased by applying additional incapable processes. Thus, the minimum amount of noise added to $\chi_{\text{expt}}\circ\chi_{\mathcal{I}n}$ should be equal to (or less than) the minimum amount of noise added to $\chi_{\text{expt}}$. In accordance with equations~(\ref{rob}) and (\ref{ROB}), we define $\beta_{n}$ as the optimal result for $\chi_{\text{expt}}\circ\chi_{\mathcal{I}n}$. We now get
\begin{equation}
\beta_{n}\leq \beta,
\end{equation}
from which we conclude that
\begin{equation}
\sum_{n}p_{n}\beta_{n}\leq\beta.
\end{equation}
In other words, the capability robustness measure possesses the property, $\sum_{n}p_{n}C(\chi_{\text{expt}}\circ\chi_{\mathcal{I}n})\leq C(\chi_{\text{expt}})$.

\noindent (MP3): The process $\chi_{\text{expt}}\circ\chi_{\mathcal{I}}$ (\ref{ten}) can be expressed as
 \begin{equation}
 \chi_{\text{expt}}\circ\chi_{\mathcal{I}}=(1+b_{\mathcal{I}})\chi'_{\mathcal{I}}-b_{\mathcal{I}}\chi'',\tag{22}
 \end{equation}
 where $\chi''=$ $\sum_{n}p_{n}\beta_{n}\chi'_{n'}/b_{\mathcal{I}},$
 $\chi'_{\mathcal{I}}=$ $\sum_{n}p_{n}(1-\beta_{n})\chi_{\mathcal{I}n'}/(1-b_{\mathcal{I}}),$ and
 $b_{\mathcal{I}}=\sum_{n}p_{n}\beta_{n}$. From the definition of the capability robustness, we know that $\beta_{\mathcal{I}}=C(\sum_{n}p_{n}\chi_{\text{expt}}\circ\chi_{\mathcal{I}n})$ and $\beta_{\mathcal{I}}\leq b_{\mathcal{I}}$. We thus show that
\begin{equation}
\beta_{\mathcal{I}}\leq \sum_{n}p_{n}\beta_{n} .\tag{23}
\end{equation}
With $\beta_{n}=C(\chi_{\text{expt}}\circ\chi_{\mathcal{I}n})$, the capability robustness measure then has the property $C(\sum_{n}p_{n}\chi_{\text{expt}}\circ\chi_{\mathcal{I}n})\leq \sum_{n}p_{n}C(\chi_{\text{expt}}\circ\chi_{\mathcal{I}n})$.

\noindent \text{(MP2)}: (2a)

Properties (2b) and (MP3) lead property (2a) as follows:
\begin{equation}
\beta_{\mathcal{I}}\stackrel{(\text{MP3})}{\leq} \sum_{n}p_{n}\beta_{n}\stackrel{(\text{2b})}\leq \beta.\tag{24}
\end{equation}
Thus, the capability robustness measure has the property $C(\chi_{\text{expt}}\circ\chi_{\mathcal{I}})\leq C(\chi_{\text{expt}})$.\\

\noindent\textbf{Comparison between $\alpha$ and $\beta$.} The idea underlying the capability composition measure highlights whether a process can be described by the model of an incapable process. Such a feature indicates that this measure is unable to reveal the distinction between processes, in general. By contrast, the idea of how close a process is to an incapable process, which underlies the capability robustness measure, makes the difference between processes visible.\\

\noindent\textbf{Constraints on incapable processes.} In this section, we present the constraint sets $D(\tilde{\chi}_{\mathcal{I}})$ required in the capability examples presented in the main text, where $\tilde{\chi}_{\mathcal{I}}$ is an unnormalized process matrix \cite{Nielsen00}. To obtain the constraints for the incapable process $D(\tilde{\chi}_{\mathcal{I}})$, we provide in the following  the limitations on both the prepared input states and their corresponding output states, which are used to construct the process matrix \cite{Nielsen00}.

\noindent{E1. Non-classical dynamics}.
To demonstrate the constraint set for the incapable processes of non-classical dynamics $D(\tilde{\chi}_{\mathcal{I},\text{ncl}})$, shown in Fig.~\ref{CZtwo}, we first introduce classical processes \cite{Hsieh17} for a two qubit system. In general, a classical process can be described by its classical states and their evolution. The classical states of input systems satisfy the assumption of realism and can be represented by the realistic sets $\textbf{v}_{\xi}\equiv(v_{1}^{1},v_{2}^{1},v_{3}^{1},v_{1}^{2},v_{2}^{2},v_{3}^{2})$, in which $v_{i}^{j}\in\{+1,-1\}$ represents the possible measurement outcomes of the $i$th physical property for the $j$th classical object. The evolutions of these states are described by transition probabilities, $\Omega_{\xi\mu}$, from $\textbf{v}_{\xi}$ to a final state $\mu$, which can be reconstructed as a density operator $\rho_\mu$ by using state tomography. Such an evolution can always be rephrased as the transition from a specific state set $\textbf{v}_{\xi}$ to some final state $\tilde{\rho}_{\mu'}$ with $\Omega_{\xi\mu'}=1$, where $\tilde{\rho}_{\mu'}$ denotes the unnormalized density matrix. The constraint set $D(\tilde{\chi}_{\mathcal{I},\text{ncl}})$ for the incapable process of non-classical dynamics is given as
\begin{equation}
\tilde{\rho}_{\mu'}\geq0\ \ \ \forall \mu', \\\tag{25}
\end{equation}
which ensures that all the output states $\tilde{\rho}_{\mu'}$ of classical processes are positive semi-definite.

The constraint set $D(\tilde{\chi}_{\mathcal{I},\text{ncl}})$ determines the number of variables that need to be optimized via SDP. For a two qubit system, each $\tilde{\rho}_{\mu'}$ in $D(\tilde{\chi}_{\mathcal{I},\text{ncl}})$ is a $4\times4$ matrix that contains 16 variables. To describe classical dynamics, we need 64 matrices which corresponds to the number of $\textbf{v}_{\xi}$, so there are 1024 variables that need to be solved by SDP.

\noindent{E2. Entanglement generation}.
To derive the constraint set $D(\tilde{\chi}_{\mathcal{I},\text{ent}})$ for entanglement generation, we first introduce the process tomography operation for a two qubit system. Any input state $\rho_{\text{in}}$ for a process can be expanded as $\rho_{\text{in}}=\sum_{k,l}\xi_{kl}\sigma_{k}\otimes \sigma_{l}$, where $\sigma_{0}=\hat{I}$ is an identity matrix; $\sigma_{1}$, $\sigma_{2}$, $\sigma_{3}$ are Pauli matrices $X$, $Y$, $Z$, respectively; and $\xi_{kl}$ is a coefficient of the form $\xi_{kl}=\text{tr}(\rho_{\text{in}}\sigma_{k}\otimes \sigma_{l})$. Similarly, the output state of the process, $\rho_{\text{out}}$, can be constructed as $\rho_{\text{out}}=\sum_{k,l}\xi_{kl}\rho_{kl}$, where $\rho_{kl}$ is the output corresponding to $\sigma_{k}\otimes \sigma_{l}$.
Since each Pauli matrix $\sigma_{1}$, $\sigma_{2}$, $\sigma_{3}$ can be represented as the spectral decomposition $\sigma_{k}=\sum_{m=\pm1}m\ket{k_{m}}\bra{k_{m}},$ where $\ket{k_{m}}$ is the eigenstate corresponding to eigenvalue $m$ of Pauli matrix $\sigma_{k}$, the input states used for process tomography are the tensor product of the eigenstates of the three Pauli matrices $\rho_{\text{in}}=\ket{k_{m}}\bra{k_{m}}\otimes\ket{l_{n}}\bra{l_{n}}$, and the corresponding output states are $\tilde{\rho}_{k_{m}l_{n}}$. The outputs of the Pauli matrices can then be represented as $\rho_{kl}=\sum_{m,n=\pm1}mn\tilde{\rho}_{k_{m}l_{n}}$.

Since all of the input states $\rho_{\text{in}}$ prepared for process tomography are separable states, the output states, $\tilde{\rho}_{k_{m}l_{n}}$, must also be separable states for a process with no ability to generate entangled states. The constraint set $D(\tilde{\chi}_{\mathcal{I},\text{ent}})$ for the incapable process of entanglement generation is given by
\begin{equation}
\begin{split}
\tilde{\rho}_{k_{m}l_{n}}\geq0,\tilde{\rho}^\text{PT}_{k_{m}l_{n}}\geq0,\ \ \ \forall k,l,m,n; \\
\sum_{m=\pm1}\tilde{\rho}_{k_{m}l_{n}}=\sum_{m=\pm1}\tilde{\rho}_{1_{m}l_{n}},\ \ \ \forall k; \\
\sum_{n=\pm1}\tilde{\rho}_{k_{m}l_{n}}=\sum_{n=\pm1}\tilde{\rho}_{k_{m}1_{n}},\ \ \ \forall l; \\ \sum_{m=\pm1}\sum_{n=\pm1}\tilde{\rho}_{k_{m}l_{n}}=\sum_{m=\pm1}\sum_{n=\pm1}\tilde{\rho}_{1_{m}1_{n}},\ \ \ \forall k,l. \label{Dchifent}
\end{split}\tag{26}
\end{equation}

The first criterion in (\ref{Dchifent}) ensures that the output states $\tilde{\rho}_{k_{m}l_{n}}$ are separable states for $\tilde{\chi}_{\mathcal{I},\text{ent}}$ by using the positive partial transpose criterion \cite{Peres96,Horodecki96}, where $\tilde{\rho}^\text{PT}_{k_{m}l_{n}}$ are the density matrices used to perform partial transpose on $\tilde{\rho}_{k_{m}l_{n}}$. The remaining constraints in (\ref{Dchifent}) ensure that when the input states are an identity matrix $\hat{I}$, the corresponding output states are the same for different decompositions of $\hat{I}$. Such a requirement is used in tomographically characterizing separable output states $\rho_{\text{out}}$ of incapable processes, where the outputs of Pauli matrices associated with the identity matrix in the inputs $\sigma_{k}\otimes \sigma_{l}$ should be objectively described, independent of measurement bases. Since the identity matrix $\hat{I}$ can be represented as the sum of two eigenstates of the Pauli matrices, i.e., $\hat{I}=\sum_{m=\pm1}\ket{k_{m}}\bra{k_{m}},  \forall k$, the output state of $\hat{I}$ for $\tilde{\chi}_{\mathcal{I},\text{ent}}$ must be the same no matter how it is decomposed. For the second constraint in equation~(\ref{Dchifent}), which states that the input of the first qubit is $\hat{I}$, the outputs $\tilde{\chi}_{\mathcal{I},\text{ent}}(\hat{I}\otimes\ket{l_{n}}\bra{l_{n}})$ are the same for all decompositions of $\hat{I}$. The third and fourth constraints in equation~(\ref{Dchifent}) state that the input of the second qubit and the input of both qubits are $\hat{I}$, respectively.

\noindent{E3. Coherence creation and
preservation}. To construct the process matrix using process tomography \cite{Nielsen00}, we first prepare the input states
\begin{equation}
\begin{split}
\ket{\psi_{kmn}} = \left\{\begin{array}{ll}
                 \frac{1}{2}(\ket{m}+\ket{n}),          & \mbox{if $k=1$, $m=n$} \\
                 \frac{1}{\sqrt{2}}(\ket{m}+\ket{n}),   & \mbox{if $k=2$, $m\neq n$} \\                \frac{1}{\sqrt{2}}(\ket{m}+i\ket{n}),  & \mbox{if $k=3$, $m\neq n$,}
                \end{array} \right.\label{StateForTomography}
\end{split}\tag{27}
\end{equation}
where $m,n=0,1,...,d-1$.
The density matrix of the input states has the form $\rho_{\text{in}}$=$\rho_{kmn}$=$\ket{\psi_{kmn}}\!\bra{\psi_{kmn}}$.
We then obtain the output states $\tilde{\chi}_{\mathcal{I}}(\rho_{kmn})$, which correspond to the prepared input states. An arbitrary output state $\tilde{\chi}_{\mathcal{I}}(\rho)$ can be obtained as a linear combination of the output states $\tilde{\chi}_{\mathcal{I}}(\rho_{kmn})$ \cite{Nielsen00}. The above specification is also applicable to superposition.

The constraint set $D(\tilde{\chi}_{\mathcal{I},\text{cre}})$ for the incapable process of coherence creation is given by
\begin{equation}
\begin{split}
\tilde{\chi}_{\mathcal{I},\text{cre}}(\rho_{kmn})&\geq0,\ \ \ \forall k,m,n;  \\
\tilde{\chi}_{\mathcal{I},\text{cre}}(\rho_{1mm})&=\sum_{j=0}^{d-1}p_j\ket{j}\!\bra{j}\in\tilde{s}_{\text{incoh}},\ \ \ \forall m,
\end{split}\tag{28}
\end{equation}
where $p_j\geq0$ are the non-negative coefficients of each pure state $\ket{j}\!\bra{j}$ and $\tilde{s}_{\text{incoh}}$ denotes the set of unnormalized incoherent states regarding the basis $\{\ket{j}\}$.

By contrast, the set of constraints for the incapable process of coherence preservation, $D(\tilde{\chi}_{\mathcal{I},\text{pre}})$, is given as
\begin{equation}
\begin{split}
&\tilde{\chi}_{\mathcal{I},\text{pre}}(\rho_{kmn})\geq0, \\ &\tilde{\chi}_{\mathcal{I},\text{pre}}(\rho_{kmn})=\sum_{j=0}^{d-1}p_j\ket{j}\!\bra{j}\in\tilde{s}_{\text{incoh}},\ \ \forall k,m,n.
\end{split}\tag{29}
\end{equation}

The number of variables that needs to be optimized by SDP is related to the output states $\tilde{\chi}_{\mathcal{I},\text{cre}}(\rho_{kmn})$ in $D(\tilde{\chi}_{\mathcal{I},\text{cre}})$. For single qubit system, each $\tilde{\chi}_{\mathcal{I},\text{cre}}(\rho_{kmn})$ is a $2\times2$ matrix that contains 4 variables. To determine the process matrix, we need four matrices, $\tilde{\chi}_{\mathcal{I},\text{cre}}(\rho_{100})$, $\tilde{\chi}_{\mathcal{I},\text{cre}}(\rho_{111})$, $\tilde{\chi}_{\mathcal{I},\text{cre}}(\rho_{201})$, and $\tilde{\chi}_{\mathcal{I},\text{cre}}(\rho_{301})$, so there are 16 variables that need to be solved by SDP. For two qubit system, each $\tilde{\chi}_{\mathcal{I},\text{cre}}(\rho_{kmn})$ is a $4\times4$ matrix that contains 16 variables. To determine the process matrix, we need 16 matrices, so there are 196 variables that need to be solved by SDP in two qubit system. This is less than the number of variables for non-classical dynamics, since they have different constraint sets $D(\tilde{\chi}_{\mathcal{I}})$.

For the quantum Ising model shown in Fig. \ref{CZtwo}, the system is a four-dimensional system with $d=4$. For $D(\tilde{\chi}_{\mathcal{I},\text{cre}})$, the set of constraints indicates that the output states of the incoherent states $\ket{j}\!\bra{j}, j\in\{0,1,2,3\}$ are incoherent states belonging to $\tilde{s}_{\text{incoh}}$, i.e., mixtures of the basis states $\sum_{j=0}^{3}p_j\ket{j}\!\bra{j}$.  Meanwhile, $D(\tilde{\chi}_{\mathcal{I},\text{pre}})$ ensures that the output states of all the input states prepared for process tomography are incoherent states belonging to $\tilde{s}_{\text{incoh}}$.\\
\noindent{E4. Superposition of quantum states}.
The superposition free states \cite{Theurer17} in $\tilde{s}_{\text{supf}}$ are mixtures of the basis states $\ket{h_{j}}\!\bra{h_{j}}$.
To make each superposition free state under $\tilde{\chi}_{\mathcal{I},\text{sup}}$ belong to $\tilde{s}_{\text{supf}}$, the constraints for the incapable process are that the output states of the superposition free states must satisfy $\tilde{\chi}_{\mathcal{I},\text{sup}}(\ket{h_{j}}\!\bra{h_{j}})$ $\in$ $\tilde{s}_{\text{supf}}$.

To construct the process matrix using process tomography, we first prepare the input states $\rho_{\text{in}}$ = $\rho_{kmn}$ =
$\ket{\psi_{kmn}}\!\bra{\psi_{kmn}}$
described in equation~(\ref{StateForTomography}) and the corresponding output states $\tilde{\chi}_{\mathcal{I},\text{supf}}(\rho_{kmn})$.

The basis states can be decomposed into a linear combination of the input states $\rho_{kmn}$. That is, $\ket{h_{j}}\!\bra{h_{j}}=\sum_{k,m,n}e_{jkmn}\rho_{kmn}$, where the coefficients $e_{jkmn}$ are as follows:
\begin{equation}
\begin{split}
e_{j2mn}=&\hspace{1ex}2\text{Re}[\text{tr}(\ket{h_{j}}\!\bra{h_{j}}\ket{m}\!\bra{n})],\hspace{1ex} m\neq n; \\
e_{j3mn}=&\hspace{1ex}2\text{Im}[\text{tr}(\ket{h_{j}}\!\bra{h_{j}}\ket{m}\!\bra{n})],\hspace{1ex} m\neq n;\\
e_{j1mn}=&\hspace{1ex}\text{tr}(\ket{h_{j}}\!\bra{h_{j}}\ket{m}\!\bra{n})-\frac{1}{2}\sum_{l=0,l\neq m}^{d-1}(e_{j2ml}+e_{j3ml}),\hspace{1ex} m=n.
\end{split}\tag{30}
\end{equation}

The output states of basis states $\ket{h_{j}}\!\bra{h_{j}}$ through $\tilde{\chi}_{\mathcal{I},\text{supf}}$ can be expressed as $\tilde{\chi}_{\mathcal{I},\text{supf}}(\ket{h_{j}}\!\bra{h_{j}})$ $=\sum_{k,m,n} e_{jkmn}\tilde{\chi}_{\mathcal{I},\text{supf}}(\rho_{kmn})$.

The constraints on the incapable process constituting $D(\tilde{\chi}_{\mathcal{I},\text{supf}})$ are given as
\begin{equation}
\begin{split}
&\tilde{\chi}_{\mathcal{I},\text{supf}}(\rho_{kmn})\geq0,\ \ \ \forall k,m,n;   \\
&\sum_{k,m,n}e_{jkmn}\tilde{\chi}_{\mathcal{I},\text{supf}}(\rho_{kmn})\\ &= \sum_{j=0}^{d-1}p_{j}\ket{h_{j}}\!\bra{h_{j}}\in\tilde{s}_{\text{supf}}, \ \  \forall \sum_{k,m,n}e_{jkmn}\rho_{kmn}=\ket{h_{j}}\!\bra{h_{j}},
\end{split}\tag{31}
\end{equation}
where $p_j$ are the non-negative coefficients of each pure state $\ket{h_{j}}\!\bra{h_{j}}$.

For the quantum Ising model in Fig.~\ref{CZtwo}, we use a four-dimensional system to describe a composite two-dimensional system, i.e., $\ket{00}=\ket{0}_{4}$, $\ket{01}=\ket{1}_{4}$, $\ket{10}=\ket{2}_{4}$ and $\ket{11}=\ket{3}_{4}$. Moreover, we choose a normalized and linear independent basis: $\ket{h_{0}}=\ket{0}_{4}$, $\ket{h_{1}}=(\ket{0}_{4}+\ket{1}_{4})/\sqrt{2}$, $\ket{h_{2}}=(\ket{0}_{4}+\ket{2}_{4})/\sqrt{2}$ and $\ket{h_{3}}=(\ket{0}_{4}+\ket{3}_{4})/\sqrt{2}$.
The superposition free states can be decomposed into the following prepared input states: $\ket{h_{0}}\!\bra{h_{0}}=e_{0100}\rho_{100}$, $\ket{h_{1}}\!\bra{h_{1}}=e_{1201}\rho_{201}$, $\ket{h_{2}}\!\bra{h_{2}}=e_{2202}\rho_{202}$, and $\ket{h_{3}}\!\bra{h_{3}}=e_{3203}\rho_{203}$, where $e_{0100}= e_{1201}=e_{2202}=e_{3203}=1$.\\

\noindent\textbf{Conversion from coherence to entanglement.} In order to evaluate the coherence conversion efficiency of process $\chi_{\text{expt}}(t)$, we choose the maximally coherent and separable state $\ket{s}=(\ket{00}+\ket{01}+\ket{10}+\ket{11})/2$ as the input state. Furthermore, we define the conversion efficiency from coherence to entanglement, ${\eta_{\text{ent}}}(t)$, as  the ratio of the concurrence \cite{Wootters98} of the output state $\chi_{\text{expt}}(t)(\ket{s}\!\!\bra{s})$, denoted as $\textit{C}_{\text{ent}}(t)$, to the normalized coherence robustness of the input state, $\tilde{s}_{\text{coh}} = 1 $. In other words, ${\eta_{\text{ent}}}(t)=\textit{C}_{\text{ent}}(t)$. (Note that, before normalization, the coherence robustness of the input state \cite{Napoli16} is $s_{\text{coh}}=3$.) For example, we have ${\eta_{\text{ent}}}(\pi/2)=0.7071$ and ${\eta_{\text{ent}}}(\pi)=1$ for a process with no depolarizing effect.

We note that these values are exactly the same as those for the entanglement generation robustness, $\beta$, for $\chi_{\text{expt}}(\pi/2)$ and $\chi_{\text{expt}}(\pi)$, respectively (see Fig.~\ref{CZtwo}b). Notably, this fact holds for all other interaction times. Moreover, for processes with a depolarizing effect, the change in conversion efficiency over time is highly coincident with the change in the entanglement generation robustness. For example, we have ${\eta_{\text{ent}}}(\pi/2)=0.6743$ and ${\eta_{\text{ent}}}(\pi)=0.9087$, which correspond to $\beta= 0.6698$ and $0.9087$, respectively. Therefore, the capability of entanglement generation can also be used to examine the coherence conversion efficiency.\\

\noindent\textbf{Conversion from coherence to superposition.} The conversion efficiency from coherence to superposition for process $\chi_{\text{expt}}(t)$ is defined as ${\eta_{\text{sup}}}(t)={\textit{s}_{\text{sup}}}/{\tilde{s}_{\text{coh}}}$, where ${\textit{s}_{\text{sup}}}$ is the normalized superposition robustness \cite{Theurer17} of the output state from $\chi_{\text{expt}}(t)$, and $\tilde{s}_{\text{coh}} = 1 $ denotes the normalized coherence robustness of the input state. To evaluate the coherence conversion efficiency, let the input state, $\ket{h_{3}}$, be chosen from the basis $\{\ket{h_{0}} = \ket{00},\ket{h_{1}} = (\ket{00}+\ket{01})/\sqrt{2},\ket{h_{2}} = (\ket{00}+\ket{10})/\sqrt{2},\ket{h_{3}} = (\ket{00}+\ket{11})/\sqrt{2}\}$. (Note that $\ket{h_{3}}$ is a coherent and superposition free state, and the coherence robustness of the input state before normalization is $s_{\text{coh}}=1$.)
For example, ${\eta_{\text{sup}}}(\pi/2)=0.621$ and ${\eta_{\text{sup}}}(\pi)=1$ correspond to $\beta= 0.8624$ and $\beta=1.366$, respectively, for $\chi_{\text{expt}}(t)$ with no depolarizing effect (see Fig.~\ref{CZtwo}b).
For $\chi_{\text{expt}}(t)$ with a depolarizing effect, ${\eta_{\text{sup}}}(\pi/2)=0.624$ and ${\eta_{\text{sup}}}(\pi)=0.9848$ correspond to $\beta= 0.8489$ and $\beta=1.361$, respectively. Notably, depolarization at $t=\pi/2$ causes the ${\eta_{\text{sup}}}(\pi/2)$ to increase, but makes the superposition robustness $\beta$ decrease.
As a result, the capability of superposition cannot be used to examine the coherence conversion efficiency.\\

\noindent\textbf{Maximum extraction of resources from processes.} Since $\chi_{\mathcal{I}}(\rho_{\text{in}})=\rho_{\mathcal{I}}$ is incapable for all incapable input states $\rho_{\text{in}}$, equations~(\ref{com}) and (\ref{rob})
 can be rephrased as $\rho_{\text{out}}=a\rho_{\mathcal{C}}+(1-a)\rho_{\mathcal{I}}$ and $(\rho_{\text{out}}+b\rho')/(1+b)=\rho_{\mathcal{I}}$, respectively, where $\rho'=\chi'(\rho_{\text{in}})$, and $\rho_{\mathcal{C}}=\chi_{C}(\rho_{\text{in}})$ are assumed to possess the prescribed resources. One can consider the minimum values of $a$ and $b$ as the quantities of resources that can be found in $\rho_{\text{out}}$ under a given input state $\rho_{\text{in}}$ and for all possible $\rho_{\mathcal{I}}$ and $\rho'$. By maximizing the quantity of  resources for all incapable input states, we obtain the maximum amount of resources that can be extracted from $\chi_{\text{expt}}$. Since the above optimization tasks have been considered in equations~(\ref{COM}) and (\ref{ROB}), we conclude that $\alpha$ and $\beta$ can quantify the maximum extraction of resources from a process.

\section*{Acknowledgements}

This work is partially supported by the Ministry of Science and Technology, Taiwan, under Grant Number MOST 107-2628-M-006-001-MY4.

\section*{Author contributions statement}

C.-M.L. devised the basic model and supervised the project. C.-C.K., S.-H.C., W.-T.L, H.L. and C.-M.L. established the final framework. C.-C.K., S.-H.C., W.-T.L, and H.-M.C. performed calculations. All authors contributed to the writing and editing of the manuscript.

%\begin{figure}[ht]
%\centering
%\includegraphics[width=\linewidth]{stream}
%\caption{Legend (350 words max). Example legend text.}
%\label{fig:stream}
%\end{figure}

%\begin{table}[ht]
%\centering
%\begin{tabular}{|l|l|l|}
%\hline
%Condition & n & p \\
%\hline
%A & 5 & 0.1 \\
%\hline
%B & 10 & 0.01 \\
%\hline
%\end{tabular}
%\caption{\label{tab:example}Legend (350 words max). Example legend text.}
%\end{table}

%Figures and tables can be referenced in LaTeX using the ref command, e.g. Figure \ref{fig:stream} and Table \ref{tab:example}.


\begin{thebibliography}{99}

\bibitem{Feynman61} Feynman, R. P. \textit{Theory of Fundamental Processes} (Westview Press, 1961).

\bibitem{Peres95} Peres, A. \textit{Quantum Theory: Concepts and Methods} (Springer, 1995).

\bibitem{Breuer&Petruccione02} Breuer, H.-P., Petruccione, F. \textit{The Theory of Open Quantum Systems} (Oxford Univ. Press, 2002).

\bibitem{Milburn96} Milburn, G. \textit{Quantum Technology} (Allen \& Unwin, 1996).

\bibitem{Feynman82} Feynman, R. P. Simulating physics with computers. \textit{Int. J. Theor. Phys.} \textbf{21}, 467–488 (1982).

\bibitem{Dowling03} Dowling, J. P. \& Milburn, G. J. Quantum technology: the second quantum revolution. \textit{Phil. Trans. A} \textbf{361}, 1655–1674 (2003).

\bibitem{O'Brien09} O’brien, J. L., Furusawa, A. \& Vu{\v{c}}kovi{\'c}, J. Photonic quantum technologies. \textit{Nat. Photonics} \textbf{3}, 687 (2009).

\bibitem{Georgescu12} Georgescu, I., Nori, F. Quantum technologies: an old new story. \textit{Physics World} \textbf{25}, 16 (2012).

\bibitem{Nielsen00} Nielsen, M.~A. \& Chuang, I.~L. \textit{Quantum Computation and Quantum Information} (Cambridge Univ. Press, 2000).

\bibitem{Ladd10} Ladd, T.~D. \textit{et~al.} Quantum computers. \textit{Nature} \textbf{464}, 45--53 (2010).

\bibitem{Gisin07} Gisin, N. \& Thew, R. Quantum communication. \textit{Nat. Photonics} \textbf{1}, 165 (2007).

\bibitem{Barenco95} Barenco, A. \textit{et~al.} Elementary gates for quantum computation. \textit{Phys. Rev. A} \textbf{52}, 3457 (1995).

\bibitem{Brennen15} Brennen, G., Giacobino, E. \& Simon, C. Focus on quantum memory. \textit{New. J. Phys.} \textbf{17}, 050201 (2015).

\bibitem{Bennett93} Bennett, C. H. \textit{et~al.} Teleporting an unknown quantum state via dual classical and Einstein-Podolsky-Rosen channels. \textit{Phys. Rev. Lett.} \textbf{70}, 1895 (1993).

\bibitem{Xia18} Xia, X.-X., Sun, Q.-C., Zhang, Q. \& Pan, J.-W. Long distance quantum teleportation. \textit{Quantum Sci. Technol.} \textbf{3}, 014012 (2018).

\bibitem{Zukowski93} Zukowski, M., Zeilinger, A., Horne, M. A. \& Ekert, A. K. “Event-ready-detectors” Bell experiment via entanglement swapping. \textit{Phys. Rev. Lett.} \textbf{71}, 4287–4290 (1993).

\bibitem{Lambert10} Lambert, N., Emary, C., Chen, Y.-N. \& Nori, F. Distinguishing quantum and classical transport through
nanostructures. \textit{Phys. Rev. Lett.} \textbf{105}, 176801 (2010).

\bibitem{Li12} Li, C.-M., Lambert, N., Chen, Y.-N., Chen, G.-Y. \&
Nori, F. Witnessing quantum coherence: from solid-state
to biological systems. \textit{Sci. Rep.} \textbf{2}, 885 (2012).

\bibitem{Emary13} Emary, C., Lambert, N. \& Nori, F. Leggett–Garg inequalities. \textit{Rep. Prog. Phys.} \textbf{77}, 016001 (2013).

\bibitem{Brunner14} Brunner, N., Cavalcanti, D., Pironio, S., Scarani, V. \& Wehner, S. Bell nonlocality. \textit{Rev. Mod. Phys.} \textbf{86}, 419 (2014).

\bibitem{Baumgratz14} Baumgratz, T., Cramer, M. \& Plenio, M. B. Quantifying coherence. \textit{Phys. Rev. Lett.} \textbf{113}, 140401 (2014).

\bibitem{Napoli16} Napoli, C. \textit{et al.} Robustness of coherence: an operational and observable measure of quantum coherence. \textit{Phys. Rev. Lett.} \textbf{116}, 150502 (2016).

\bibitem{Winter16} Winter, A. \& Yang, D. Operational resource theory of coherence. \textit{Phys. Rev. Lett.} \textbf{116}, 120404 (2016).

\bibitem{Theurer17} Theurer, T., Killoran, N., Egloff, D. \& Plenio, M. B. Resource theory of superposition. \textit{Phys. Rev. Lett.} \textbf{119}, 230401 (2017).

\bibitem{Vogel89} Vogel, K. \& Risken, H. Determination of quasiprobability distributions in terms of probability distributions for the rotated quadrature phase. \textit{Phys. Rev. A} \textbf{40}, 2847 (1989).

\bibitem{Leonhardt97} Leonhardt, U. \textit{Measuring the Quantum State of Light} (Cambridge Univ. Press, Cambridge, England, 1997).

\bibitem {Chuang97}Chuang, I. L. \& Nielsen, M. A. Prescription for experimental determination of the dynamics of a quantum black box. \textit{J. Mod. Opt.} \textbf{44}, 2455–2467 (1997).

%\bibitem{Lofberg}
%\bibinfo{author}{Lofberg, J.}
%\newblock \bibinfo{title}{\text{YALMIP}: A toolbox for modeling and
%  optimization in \text{MATLAB}}.
%\newblock In \emph{\bibinfo{booktitle}{CACSD, 2004 IEEE International Symposium
%  on Taipei, Taiwan}}, \bibinfo{pages}{Available:
%  \url{http://users.isy.liu.se/johanl/yalmip/}.}

\bibitem{Lofberg} L{\"o}fberg, J. Yalmip: A toolbox for modeling and optimization in MATLAB. \textit{In CACSD, 2004 IEEE International Symposium on Taipei, Taiwan)}. Available at : http://users.isy.liu.se/johanl/yalmip/.

\bibitem{sdpsolver} Toh, K. C., Todd, M. J. \& Tütüncü, R. H. SDPT3 -- a MATLAB software package for semidefinite-quadratic-linear programming, version 4.0. Available at :
https://github.com/sqlp/sdpt3.

\bibitem{Hsieh17} Hsieh, J.-H., Chen, S.-H. \& Li, C.-M. Quantifying quantum-mechanical processes. \textit{Sci. Rep.} \textbf{7}, 13588 (2017).

\bibitem{Pan12} Pan, J.-W. \textit{et al.} Multiphoton entanglement and interferometry. \textit{Rev. Mod. Phys.} \textbf{84}, 777 (2012).

\bibitem{Peres96} Peres, A. Separability criterion for density matrices. \textit{Phys. Rev. Lett.} \textbf{77}, 1413 (1996).

\bibitem{Horodecki96} Horodecki, M., Horodecki, P. \& Horodecki, R. Separabiliy of mixed states: Necessary and sufficient conditions. \textit{Phys. Lett. A} \textbf{223}, 1 (1996).

\bibitem{Dirac30} Dirac, P. A. M. \textit{The Principles of Quantum Mechanics,} (The Clarendon Press, 1930).

\bibitem{Briegel01} Briegel, H. J. \& Raussendorf, R. Persistent entanglement in arrays of interacting particles. \textit{Phys. Rev. Lett.} \textbf{86}, 910 (2001).

\bibitem{Raussendorf01} Raussendorf, R. \& Briegel, H. J. A one-way quantum computer. \textit{Phys. Rev. Lett.} \textbf{86}, 5188 (2001).

\bibitem{Yuen-Zhou11} Yuen-Zhou, J., Krich, J. J., Mohseni, M. \& Aspuru-Guzik, A. Quantum state and process tomography of energy transfer systems via ultrafast spectroscopy. \textit{Proc. Natl. Acad. Sci. U.S.A.} \textbf{108}, 17615–17620 (2011).

\bibitem{You11} You, J. \& Nori, F. Atomic physics and quantum optics using superconducting circuits. \textit{Nature} \textbf{474}, 589 (2011).

\bibitem{Guhne09} G{\"u}hne, O. \& T{\'o}th, G. Entanglement detection. \textit{Phys. Rep.} \textbf{474}, 1–75 (2009).

\bibitem{Wiseman07} Wiseman, H. M., Jones, S. J. \& Doherty, A. C. Steering, entanglement, nonlocality, and the Einstein-Podolsky-Rosen paradox. \textit{Phys. Rev. Lett.} \textbf{98}, 140402 (2007).

\bibitem{He13} He, Q. \& Reid, M. Genuine multipartite Einstein-Podolsky-Rosen steering. \textit{Phys. Rev. Lett.} \textbf{111}, 250403 (2013).

\bibitem{Li15} Li, C.-M. \textit{et al.} Genuine high-order Einstein-Podolsky-Rosen steering. \textit{Phys. Rev. Lett.} \textbf{115}, 010402 (2015).

\bibitem{Wang18} Wang, K.-H. \textit{et al.} Non-Markovianity of photon dynamics in a birefringent crystal. \textit{Phys. Rev. A} \textbf{98}, 043850 (2018).

\bibitem{Vega17} De Vega, I. \& Alonso, D. Dynamics of non-markovian open quantum systems. \textit{Rev. Mod. Phys.} \textbf{89}, 015001 (2017).

\bibitem{Vedral98} Vedral, V. \& Plenio, M. B. Entanglement measures and purification procedures. \textit{Phys. Rev. A} \textbf{57}, 1619 (1998).

\bibitem{Plenio07} Plenio, M. B. \& Virmani, S. An introduction to entanglement measures. \textit{Quantum Info. Comput.} \textbf{7}, 1–51 (2007).

\bibitem{Shabani11} Shabani, A. \textit{et al.} Efficient measurement of quantum dynamics via compressive sensing. \textit{Phys. Rev. Lett.} \textbf{106}, 100401 (2011).

\bibitem{Johansson14} Johansson, J. R., Lambert, N., Mahboob, I., Yamaguchi, H. \& Nori, F. Entangled-state generation and Bell inequality violations in nanomechanical resonators. \textit{Phys. Rev. B} \textbf{90}, 174307 (2014).

\bibitem{Wootters98} Wootters, W. K. Entanglement of formation of an arbitrary state of two qubits. \textit{Phys. Rev. Lett.} \textbf{80}, 2245 (1998).
\end{thebibliography}
\end{document}